# Polymer-Nanoparticle Complexes : from Dilute Solution to Solid State

**Jean-François Berret**[@]
Matière et Systèmes Complexes, UMR CNRS n° 7057,
Université Denis Diderot Paris-VII, 140 rue de Lourmel 75015 Paris, France

**Kazuhiko Yokota**
Rhodia, Centre de Recherches d'Aubervilliers,
52 rue de la Haie Coq, 93308 Aubervilliers Cedex, France

**Mikel Morvan**
Complex Fluids Laboratory, UMR CNRS - Rhodia 166,
Cranbury Research Center Rhodia, 259 Prospect Plains Road, Cranbury NJ 08512, USA

**and Ralf Schweins**
Institute Laue-Langevin, LSS Group, BP 156, F-38042 Grenoble cedex 9, FRANCE

| | |
|---|---|
| *Corresponding author:* | Jean-François Berret |
| Email | jean-francois.berret@paris7.jussieu.fr |
| tel | 33 (0)1 44 27 44 97 |
| fax | 33 (0)1 44 27 38 82 |

**Abstract :** We report on the formation and the structural properties of "supermicellar" aggregates also called electrostatic complexes, made from mineral nanoparticles and polyelectrolyte-neutral block copolymers in aqueous solutions. The mineral particles put under scrutiny are ultra-fine and positively charged yttrium hydroxyacetate nanoparticles. Combining light, neutron and x-ray scattering experiments, we have characterized the sizes and the aggregation numbers of the organic-inorganic complexes. We have found that the hybrid aggregates have typical sizes in the range 100 nm and exhibit a remarkable colloidal stability with respect to ionic strength and concentration variations. Solid films with thicknesses up to several hundreds of micrometers were cast from solutions, resulting in a bulk polymer matrix in which nanoparticle clusters are dispersed and immobilized. It was found in addition that the structure of the complexes remains practically unchanged during film casting.

# I - Introduction

Development of molecular architectures is one of the final goals of modern chemistry, physics and biology. The emergence of novel materials and processing at the nano scale has enabled the synthesis of new materials, as well as it has provided decisive breakthroughs in the field of the nanotechnology[1-4]. This is the case of magnetic and luminescent nanocrystals which physical properties are nowadays exploited for imaging complex fluids at mesoscopic scales. In biomedicine, the particles can also be functionalized with ligands, peptides, oligonucleotides to reach a target or to deliver a drug to a specific location[5-7]. One of the most promising approach for the stabilization of nanoparticles in different environments, such as in solutions, thin or thick films is based on the use of polymers or of copolymers. In solutions, the first objective is to build around the particle a diffuse and protective corona. This corona aims to promote purely steric repulsions between the colloids, reducing at the same time the range and strength of the electrostatic and van der Waals interactions[8-10]. Nowadays, polymer-coated nanoparticles can be obtained by *in situ* polymerization[8,9,11] or by encapsulation into block copolymer micelles[12-17]. Polymer coatings can be also achieved by simple physical-chemistry techniques based on electrostatic complexation between oppositely charged particles and polymers. Two protocols have been designed and reported so far. In the first one, polyelectrolyte-neutral block copolymers were associated in aqueous solutions with oppositely charged surfactants or nanoparticles, yielding the formation of stable and monodisperse "supermicellar" aggregates with core-shell structures[18-21]. In a second protocol, a two-step mechanism based on precipitation-redispersion of the mixed solutions was developed[22]. As outcome of this second process, the particles are irreversibly coated with short homopolyelectrolyte chains, resulting in an increased stability for the solutions.

In the present paper, we re-examine the collective complexation investigated using neutral-charged diblocks and we apply it to yttrium hydroxyacetate (YHA) nanoparticles[18,19]. The YHA particles were chosen as a model system for inorganic colloids. Their typical sizes (~ 2 nm) and charges compare well with those of cationic surfactant micelles, a colloidal system that was thoroughly investigated with respect to complexation during the last decade[23-28]. The YHA nanoparticles were also considered because of their potential applications as precursors for ceramic and opto-electronic materials[29]. In our earlier accounts on the YHA-diblock mixed systems, we had demonstrated the analogy between surfactant micelles and nanoparticles in the complexation phenomenon[18,19]. In the present report, we stress the similarities between the nanostructures formed by spontaneous electrostatic complexation and the polymeric micelles built from amphiphilic copolymers[30]. We are taking advantage of the enhanced stability of the YHA-diblock complexes to cast micrometer thick films from the solutions, as this is commonly done with copolymers in selective solvent[31-34]. The result of the film casting is a nanocomposite material made from clusters of nanoparticles dispersed in a polymer matrix.

# II - Experimental

## II. 1 – Characterization, Sample Preparation and Phase Diagram

*Polyelectrolyte-Neutral Diblock Copolymer* : The poly(acrylic acid)-*b*-poly(acrylamide) block copolymers used to complex YHA-nanoparticles were synthesized by controlled radical polymerization[35,36]. In the present study, we focus on the two polymers, PANa(5K)-*b*-PAM(30K) and PANa(5K)-*b*-PAM(60K), where PANa stands for poly(sodium acrylate) and PAM for poly(acrylamide). The values in parenthesis are the molecular weights targeted by the synthesis of the acid forms. Although the actual values are slightly different (see below), they will be used throughout the paper. In aqueous solutions, the chains are well dispersed and in the state of unimers (pH > 3). The weight-averaged molecular weights $M_w^{pol}$ and the hydrodynamic radius $R_H^{pol}$ were determined by static and dynamic light scattering (Brookhaven BI-9000AT spectrometer). We found $M_w^{pol}$ = 43 500 g·mol$^{-1}$, $R_H^{pol}$ = 5.5 nm for PANa(5K)-*b*-PAM(30K) and $M_w^{pol}$ = 68 300 g·mol$^{-1}$, $R_H^{pol}$ 7.9 nm for PANa(5K)-*b*-PAM(60K). The polydispersity index was determined by size exclusion chromatography at 1.6. Poly(acrylic acid) is a weak polyelectrolyte and its ionization state depends on the pH. In order to derive the molecular weight of the acrylic acid block, titration experiments were performed. By slow addition of sodium hydroxide 0.1 N, the pH of PANa-*b*-PAM solutions was varied from acidic to basic conditions. The determination of the titration curves and equivalences allowed us to get the degree of polymerization for the anionic block. We obtained $n_{PE}$ = 90 for PANa(5K)-*b*-PAM(30K) and $n_{PE}$ = 77 for PANa(5K)-*b*-PAM(60K), yielding molecular weights of 6500 and 5560 g·mol$^{-1}$, respectively. Supplementary information for these polymers are available in Ref.[28].

*Inorganic Nanoparticles* : The synthesis of the yttrium hydroxyacetate (YHA) nanoparticles is based on a reaction described in Refs[18,19]. By the end of the reaction, the average composition of the particles is $Y(OH)_{1.7}(CH_3COO)_{1.3}$. As for the polymers, the molecular weight $M_w^{nano}$ (= 27 000 g·mol$^{-1}$) and the hydrodynamic radius $R_H^{nano}$ (~ 2 nm) of YHA nanoparticles were determined by light scattering. Zeta potential measurements (Zetasizer 3000 from Malvern) corroborates that the particles are positively charged ($\zeta$ = + 45 mV) and that the suspensions are stabilized by surface charges. The stability of $Y(OH)_{1.7}(CH_3COO)_{1.3}$ sols at neutral pH is excellent over a period of several weeks.

*Sample Preparation and Phase Diagram* : The mixed aggregates were obtained by mixing nanoparticle and polymer solutions prepared at the same weight concentration c (c ~ 1 wt. %) and at the same pH (pH 7). The mixing ratio X is defined as the volume of yttrium-based suspension relative to that of the polymer. According to the above definitions, the concentrations in nanoparticles and polymers in the mixed solutions are $c_{pol}$ = c/(1+X) and $c_{nano}$ = Xc/(1+X). In earlier reports on nanoparticle-polymer complexes, we have shown that there exists a preferred mixing ratio noted $X_P$ at which all the polymers and nanoparticles brought by mixing are involved in the formation of the hybrid colloids. For PANa-*b*-PAM/YHA complexes, $X_P$ is around 0.2[19].

Fig. 1 shows a schematic phase diagram for the ternary system comprising copolymers, nanoparticles and solvent. Solutions obtained according to the mixing procedure described above are located on the X-line. On the X-line, X varies from X = 0 to X = ∞, with the total concentration being constant. On the other hand, the c-line describes solutions at a same X, but with different concentrations. In this work, we have explored ranges in mixing ratio X varying from 0 to ∞ and in concentration c = 0.1 wt. % to ~ 100 wt. % (solid state). The polymer-nanoparticle complexes were prepared at low concentrations, usually c = 0.1 - 1 wt. % and then concentrated by solvent evaporation or by ultrafiltration. The ultrafiltration cell was equipped with a 3000 g·mol$^{-1}$ pore size filter (Pall Life Sciences). With increasing concentrations, the mixed systems pass from Newtonian liquids (c < 30 wt. %) to viscoelastic gels (c > 30 wt. %) and to solid. For the casting, open teflon cells designed to contain up to 2×2×0.5 cm$^3$ of liquid were filled with the solutions and stored at 60 °C during one day and under vacuum (0.01 atm.). These conditions allowed for a slow evaporation of the solvent. Thermal Gravimetry Analysis (TGA) performed on the films have revealed the presence of unbound water molecules of nearly 10 % by weight, *i.e.* the solid films have an actual concentrations of ~ 90 wt. %. .

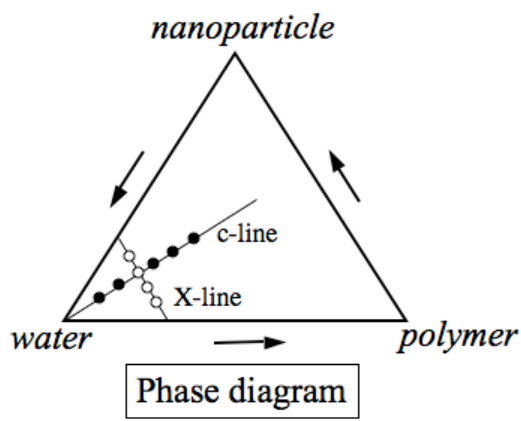

**Figure 1 :** Schematic phase diagram for the ternary system comprising copolymers, nanoparticles and water. The mixing procedures developed here provides solutions located on the X- and on the c-lines, X being the mixing ratio and c the total concentration. Here, we explore ranges in mixing ratios X = 0 – ∞ and in concentrations c = 0.1 – 90 wt. %.

## II. 2 – Small-Angle Scattering

Static and dynamic light scattering were performed on a Brookhaven spectrometer for measurements of the Rayleigh ratio (absolute scattering intensity) and of the collective diffusion constant using an experimental set-up described previously[19]. Light scattering was used to determine the apparent molecular weight $M_{w,app}$ of polymers, nanoparticles and hybrids, as well as the radius of the gyration $R_G$ of the supermicellar aggregates. From the value of diffusion constant extrapolated at c = 0, the hydrodynamic radius $R_H$ was also calculated according to the Stokes-Einstein relation. We recall that for monodisperse and homogeneous spheres of radius R, one has R = $R_H$ = 1.29 $R_G$. For core-shell microstructures e.g. obtained from amphiphilic block copolymers[30,37,38], the ratio $R_H/R_G$ has been found experimentally in the range 1 – 2. There, the hydrodynamic radius $R_H$ is assumed to be the overall radius of block copolymer micelle[39] and the shell thickness becomes $h = R_H - R_C$ where $R_C$ is the radius of the core. Note that in general the determination of the core radius can not be achieved by light scattering but requires additional techniques, such as cryogenic electron microscopy.

Small-angle x-ray scattering (SAXS) runs were performed at the Brookhaven National Laboratory (Brookhaven, USA) on the X21 and X3A2 beam lines. For the runs on X21, we used a wavelength $\lambda = 1.76$ Å for the incoming beam and the sample-detector distance of 1 meter. With the detector in the off-center position, the accessible q-range was 0.01 – 0.4 Å$^{-1}$, with a resolution $\Delta q_{FWHM}/q$ of 1.25 %. $\Delta q_{FWHM}$ denotes here the full width at half maximum of diffraction peaks characterizing ordered structures. For the run on X3A2, the detector was located at 1.5 meter, with $\lambda = 1.55$ Å and a q-range of interest 0.008 – 0.27 Å$^{-1}$. The data were treated so as to remove the contribution of the glass capillary (diameter 1.5 mm) and that of the solvent. On the two beam lines, the wave-vector scale was calibrated against silver behenate powder. Thick films casted from the mixed polymer-nanoparticle solutions (thickness 0.1 – 1 mm) were directly taped on the sample holder.

Small-angle neutron scattering was performed at the Institute Laue-Langevin (Grenoble, France) on the D11 beam line. For SANS, polymer-nanoparticle hybrids were prepared using $D_2O$ as a solvent for contrast reasons. On D11, the data collected at 1.1, 5 and 20 meters cover a range in wave-vector : $2 \times 10^{-3}$ Å$^{-1}$ to 0.35 Å$^{-1}$, with an incident wavelength of 8 Å and a wave-vector resolution $\Delta q/q$ of 10 %. The spectra are treated according to the standard Institute Laue-Langevin procedures, and the scattering cross sections are expressed in cm$^{-1}$.

The molecular weight, volumes and scattering length densities of the chemical species studied in this work are listed in Table I. The data for the YHA nanoparticles were determined by contrast variation using the neutron technique. The stock solution prepared in $H_2O$ at c = 25.5 wt. % was diluted with $D_2O$ down to a concentration of 1 wt. %. Doing so, the scattering length density of the solvent $\rho_S$ was allowed to change from $\rho_S = -0.56 \cdot 10^{10}$ cm$^{-2}$ to $6.2 \cdot 10^{10}$ cm$^{-2}$. For $\rho_S = \rho_{YHA}$, it is predicted that the scattering intensity extrapolated to zero wave-vector should be zero[40]. Fig. 2 displays the square root of the scattering intensity extrapolated at zero wave-vector and divided by the concentration for the series described previously, $\left[\frac{1}{c} d\sigma(q \rightarrow 0)/d\Omega\right]^{1/2}$. For the $H_2O$ rich mixtures, we also reversed the sign of the square root in order to emphasize the linear dependence as function of $\rho_S$. $\left[\frac{1}{c} d\sigma(q \rightarrow 0)/d\Omega\right]^{1/2}$ vanishes at $\rho_N^{nano} = 3.50 \cdot 10^{10}$ cm$^{-2}$, which corresponds the value of the average scattering length density for the nanoparticles. From $\rho_N^{nano}$, we also estimate the molar volume ($V_{mol}$ = 50.0 cm$^3$·mol$^{-1}$) and x-ray scattering length density ($\rho_X^{nano}$ = 31.9·10$^{10}$ cm$^{-2}$) for the nanocolloids, using the expressions $\rho_{N,X} = b_{N,X} N_A / V_{mol}$. Here, $b_N$ and $b_X = Z r_{el}$ are the neutron and x-ray scattering lengths of the elementary scatterer (Z denotes the number of electrons of this scatterer, $N_A$ the Avogadro number and $r_{el}$ the electron radius)[40]. As evidenced in Table I, the YHA nanoparticles and the polymers have comparable neutron scattering densities with respect to D$_2$O, and thus both will contribute to the SANS cross-sections. By SAXS, the contrast of the inorganic part is much larger than that of the polymers, and as a result the nanoparticles will mainly dominate in x-ray.

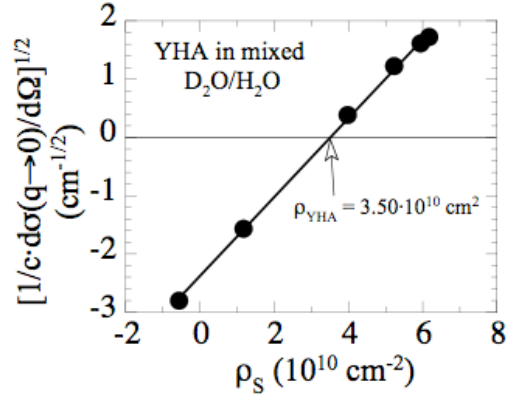

**Figure 2 :** Square root of the SANS scattering intensity extrapolated at zero wave-vector, $[\frac{1}{c} d\sigma(q \rightarrow 0)/d\Omega]^{1/2}$ for a series of dilute solutions made with mixed H$_2$O/D$_2$O solvent. The scattering length density of the solvent is allowed to change from $\rho_S$ = -0.56·10$^{10}$ cm$^{-2}$ to 6.2·10$^{10}$ cm$^{-2}$. $[\frac{1}{c} d\sigma(q \rightarrow 0)/d\Omega]^{1/2}$ cancels at $\rho_N^{nano}$ = 3.50·10$^{10}$ cm$^{-2}$, which is the value of the average scattering length density of the bare nanoparticles. Note that for the H$_2$O rich mixtures we have reversed the sign of the square root in order to emphasize the linear dependence as function of $\rho_S$.

## II. 3 – Cryo-Transmission Electron Microscopy

Cryo-transmission electron microscopy (cryo-TEM) experiments were performed on polymer-nanoparticle solutions made at the preferred ratio and concentration c = 0.2 wt. %. For cryo-TEM experiments, a drop of the solution was put on a TEM-grid covered by a 100 nm-thick polymer perforated membrane. The drop was blotted with filter paper and the grid was quenched rapidly in liquid ethane in order to avoid the crystallization of the aqueous phase. The membrane was finally transferred into the vacuum column of a TEM-microscope (JEOL 1200 EX operating at 120 kV) where it was maintained at liquid nitrogen temperature. The magnification for the cryo-TEM experiments was selected at 20 000×.

# III – Results and Discussion

## III.1 – Small-Angle X-Ray Scattering

Fig. 3 illustrates the SAXS intensities obtained at c = 1 wt. % and X varying between X = 0 and X = ∞ (Fig. 1). For the pure polymer solution (X = 0), the overall intensity is weak and it follows at high q a scaling law with an exponent of –1.4[40,41]. The data for X = ∞ represent the form factor of the YHA-nanoparticles, obtained at slightly lower concentration (c = 0.4 wt. %).

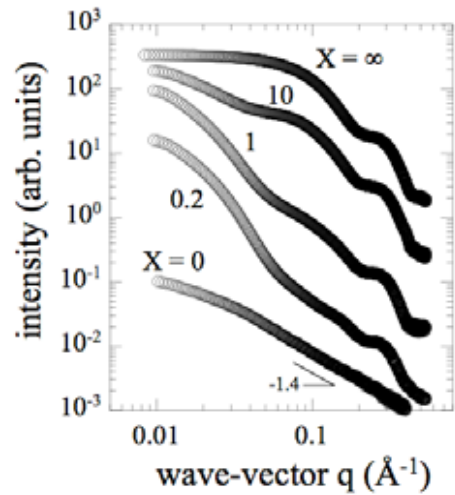

**Figure 3 :** Small-angle x-ray scattering intensities obtained for mixed nanoparticle-polymer solutions at c = 1 wt. % and X comprised between 0 and 10. Each curve has been shifted for clarity. The upper curve for X = ∞ is the form factor of the YHA nanoparticles[18].

The YHA form factor in Fig. 3 exhibits a two-step decrease. The first decrease occurring at low q is associated with the overall size of the particle and provides a radius of gyration of $R_G^{nano}$ = 1.65 nm. At higher q, a shoulder shows up around 0.3 Å$^{-1}$ which is the signature of a complex internal structure and which involves much likely two layers with different scattering length densities. For the intermediate X-values, X = 0.2, 1 and X = 10, the scattering cross-section is dominated by a strong forward scattering (*i.e.* as q → 0). The forward scattering is indicative for the formation of large-scale aggregates. The radius of gyration associated with the intensity decrease at low q results in $R_G$ = 11 nm, *i.e.* a value larger than that of the single nanoparticles by a factor ~ 5 (Table I). From the X-dependence of the intensity, we confirm that the preferred mixing ratio for the YHA–diblock system is $X_P \sim 0.2$. We recall here that $X_P$ is the ratio for which all the species present associate, yielding a maximum of the ($q \to 0$)-cross-section. Taking into account the molecular weight of the two components, $X_P = 0.2$ corresponds to ~ 4 polymers per particle for PANa(5K)-*b*-PAM(30K) and to ~ 2 polymers per particle for PANa(5K)-*b*-PAM(60K). In the sequel of the paper, we focus on this unique composition. Note finally the similarities in the q-dependencies between the data at $X = 10$ and $X = \infty$, especially at large wave-vectors. This shows that at high X there is a coexistence between unassociated nanoparticles and mixed aggregates.

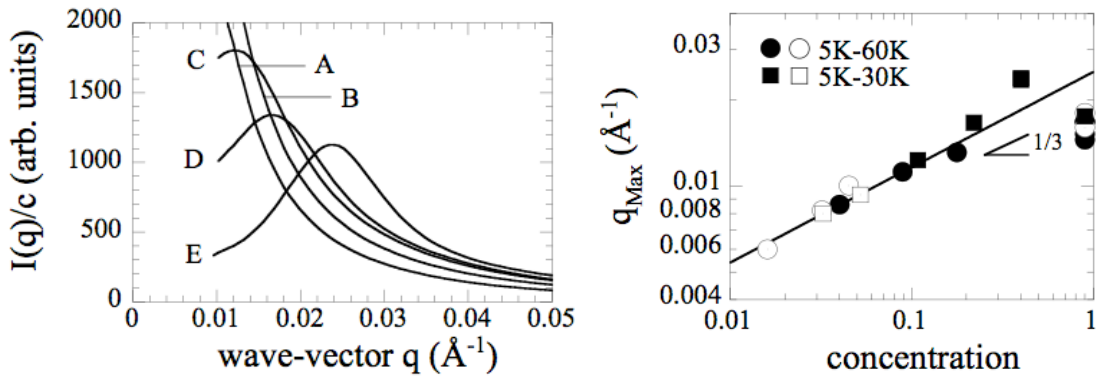

**Figure 4 :** Normalized SAXS intensities for nanoparticle-polymer solutions at the concentrations c = 2 (curve A), 3.7 (B), 10.8 (C), 22 (D) and 40 wt. % (E). The samples were obtained by slow evaporation of the solvent, starting with a solution prepared at c = 2 wt. % and using PANa(5K)-*b*-PAM(30K) as complexing polymer. The samples at c = 10.8 and 22 wt. % are viscous liquids and that at 40 wt. % is a gel.
**Figure 5 :** Evolution of the position of the structure peak $q_{Max}(c)$ in a double logarithmic representation for the PANa(5K)-*b*-PAM(30K)/YHA (squares) and PANa(5K)-*b*-PAM(60K)/YHA (circles) mixed systems as a function of the concentration. The close (open) symbols refer to SAXS (SANS) experiments. $q_{Max}(c)$ exhibits a power law with an exponent close to 1/3 in both cases.

Fig. 4 shows the normalized intensities for mixed solutions at c = 2, 3.7, 10.8, 22 and 40 wt. %. For these concentrations, the curves are labeled A, B, C, D and E respectively. The samples were obtained by slow evaporation of the solvent, starting from a solution prepared at c = 2 wt. % and using PANa(5K)-*b*-PAM(30K) as complexing agent. It is important to note here that the solutions at c = 10.8 and 22 wt. % exhibit an increased viscosity as compared to that of the solvent. These two fluids remain however Newtonian. The system at c = 40 wt. % on the contrary has the rheological properties of a strong gel[42]. The origin of the gel-like properties at high concentrations is discussed in the conclusion section. With increasing concentration in Fig. 4, there is the development and growth of a structure peak which is the signature of the strong repulsive interactions between the hybrid aggregates. Located at a wave-vector $q_{Max}(c)$, the structure peak increases with increasing concentration. The quantity $2\pi/q_{Max}(c)$ is an estimate of the d-spacing between aggregates. Fig. 5 displays the evolution of the position of the structure peak in a double logarithmic representation as a function of the mass fraction. For the two systems PANa(5K)-*b*-PAM(30K)/YHA and PANa(5K)-*b*-PAM(60K)/YHA studied, $q_{Max}(c)$ exhibits a power law with an exponent close to 1/3. This power law is characteristic for strongly interacting spherical colloids. It is an indication too that the mixed aggregates preserved their structures as c is increased.

## III.2 – Small-Angle Neutron Scattering

Fig. 6 and 7 show the scattering cross-sections obtained by SANS on PANa(5K)-*b*-PAM(30K)/YHA and PANa(5K)-*b*-PAM(60K)/YHA solutions at concentrations comprised between 0.5 and 5 wt. %. The intensities have been divided by the concentration in order to emphasize the superposition of the data in the high q-range. This approach is required to separate the contributions between the form and the structure factors of the hybrid aggregates. The intensities at low concentration, here c = 0.5 wt. % are proportional to the form factor (see insets in Figs. 6 and 7). As for the SAXS data, this form factor shows a significant forward scattering. However, at high wave-vectors, it is followed by a decrease in the form of power law. This latter regime is different from that observed in SAXS (Fig. 3) and it is due to the different scattering contrasts for the core and for the shell with respect to the two techniques. The asymptotic behaviors for the form factors are similar to those found in polymer stars, especially with polymer stars with a large number of arms[43,44]. Following Grest *et al.*[44,45], the scattering intensity for star-like objects can be approximated by:

$$\frac{d\sigma}{d\Omega}(q, c \ll 1) = \alpha \exp(-\frac{q^2 R_G^2}{3}) + \beta \frac{\sin(\mu \tan^{-1}(q\xi))}{q\xi(1+q^2\xi^2)^{\mu/2}} \quad (1)$$

The first term in the right hand side of Eq. 1 accounts for the overall size of the colloidal particle and determines its gyration radius $R_G$. The second term is the Fourier transform of the monomer-monomer correlation function within the corona in agreement with the Daoud and Cotton Model[46]. In Eq. 1, $\alpha$ and $\beta$ are prefactors, $\xi$ is the average blob size in the corona. Effective at large q, the second contribution decreases as $q^{-\mu-1}$. For Gaussian chains, $\mu = 1$ and for chains in good solvent conditions, $\mu = 2/3$. As shown in the insets, the overall q-dependence of the intensity is well accounted for by Eq. 1. Adjustable parameters are α, $R_G$, β, ξ and μ. We obtained radii of gyration $R_G$ = 22.7 nm and 18.2 nm, a power law exponent of -2 in the high q-region (*i.e.* μ = 1 in Eq. 1) and ξ = 15 - 20 nm. In Figs. 6 and 7, the superimposition of the intensities in the range 0.01 Å$^{-1}$ - 0.4 Å$^{-1}$ suggests that the internal structure of the aggregates is not altered by the removal of the solvent, and the subsequent increase of the weight and volume concentration.

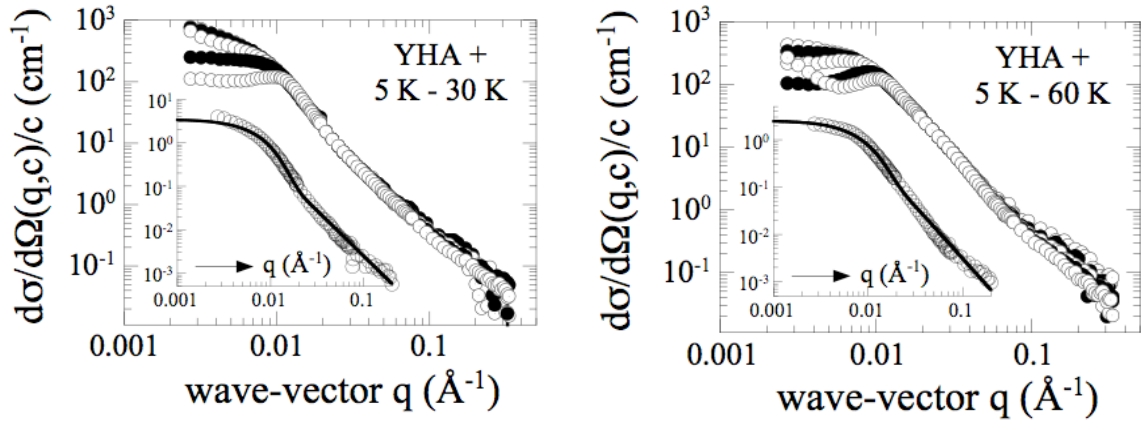

**Figure 6 :** SANS scattering cross-sections obtained on PANa(5K)-*b*-PAM(30K)/YHA solutions at concentrations comprised between 0.5 and 5 wt. %. The intensities have been divided by the concentration in order to emphasize the superposition of the data in the high q-range. Inset : neutron scattering absolute intensity (cm$^{-1}$) of the mixed aggregates obtained at c = 0.5 wt. %. As explained in the text, this intensity is proportional to the form factor of the hybrid aggregates. It is fitted using an analytical expression developed for star-like objects, provided by Eq. 1 and shown as a continuous line [44].
**Figure 7 :** Same as Fig. 6, but for PANa(5K)-*b*-PAM(60K)/YHA mixed systems.

The shift of the structure factor to higher wave-vectors is observed for the two systems with increasing concentration. The $q_{Max}(c)$-values obtained by SANS have been included in Fig. 5, and they are found to align with the SAXS data. The straight line in Fig. 5 is obtained using the expression $q_{Max}(c) = q_0 c^{1/3}$, with $q_0$ = 0.025 Å$^{-1}$. The two systems, PANa(5K)-*b*-PAM(30K)/YHA and PANa(5K)-*b*-PAM(60K)/YHA follow nearly the same behavior, a finding which is in agreement with the fact that both types of aggregates have similar hydrodynamic and gyration radii (Table II).

III.3 – Association Morphology between Polymers and Nanoparticles

Fig. 8a displays a picture obtained by cryo-TEM of the PANa(5K)-*b*-PAM(60K)/YHA solution at c = 1 wt. % and X = 0.2. Prior to the cryo-TEM experiments, the solution was characterized by static and dynamic light scattering, yielding for the gyration and hydrodynamic radii $R_G$ = 17.2 nm and $R_H$ = 33.5 nm (see Table II). The field covered on the viewgraph is ~ 600×600 nm² and was obtained with a 20 000× magnification. This magnification allows to clearly distinguish dark patches on a gray background, with radii comprised between 5 and 15 nm.

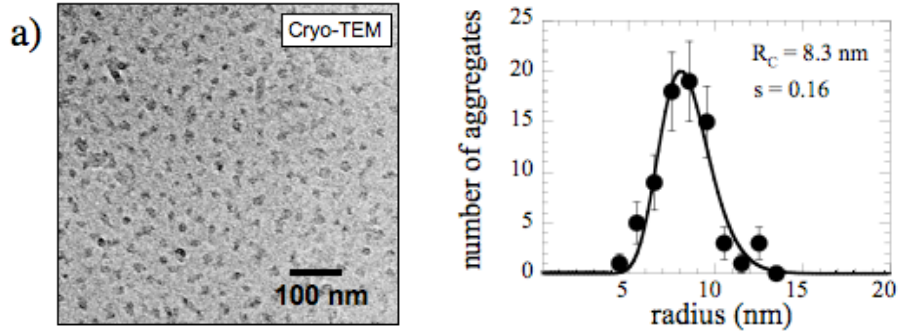

**Figure 8 :** a) Cryo-TEM image of mixed aggregates obtained by complexation between PANa(5K)-*b*-PAM(60K) and YHA. The total concentration is c = 1 wt. % and X = $X_P$ (= 0.2). The field covered on the viewgraph is ~ 600×600 nm². The patches delimitate the contours of the cores, as discussed in the text. b) Size distribution of the dark patches identified in a). The continuous line results from a fitting procedure using a log-normal function with average radius $R_C$ = 8.3 ± 0.2 nm and polydispersity s = 0.16 ± 0.02.

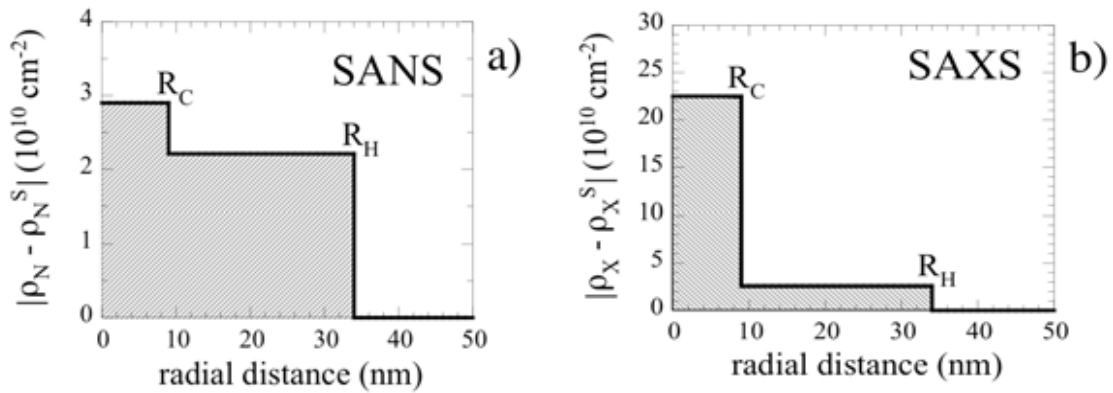

**Figure 9 :** SANS (a) and SAXS (b) contrasts of the YHA-copolymer supermicellar aggregates as determined from the comparison between scattering techniques and cryo-TEM.

An image analysis on 74 patches provides the size distribution of these patches (Fig. 8b). The distribution is found to be described by a log-normal function with an average radius $R_C$ = 8.3 ± 0.2 nm and a polydispersity s = 0.16 ± 0.02. Since the YHA nanoparticles have the largest electronic density from all the species present, we can assume that the patches in Fig. 8a delimitate the cores of the hybrid aggregates (polymers are not visible with this technique). That the cores are close to each other comes from the fact that the colloids are located at different heights within the ~ 200 nm-thick film studied by cryo-TEM. From the value of $R_C$ = 8.3 nm, we can derive two quantities which are relevant for the complexes : the thickness of the neutral polymer brush surrounding the cores, $h = R_H - R_C$ = 25 nm, and the aggregation numbers, expressed here in terms of the number of nanoparticles per aggregate. For this latter determination, the expression $N_{Agg} = \phi_C (R_C/R)^3$ is used[26,28] where $\phi_C$ (= 0.4) is the volume fraction of particles in the core and R the radius of a single particles. We obtain $N_{Agg}$ = 92, in good agreement with former determinations[18,19]. Figs 9a and 9b illustrate respectively the SANS and SAXS contrasts for core-shell aggregates with core radius and shell thickness as determined from the previous analysis.

III.4 – Thick Films Casted from Solutions

Solid thick films made from PANa-*b*-PAM/YHA were casted from dilute solutions in a two step process. In the first step, an ultrafiltration cell allows to reach a range around 10 wt. %, where the liquid becomes noticeably viscous. Then, the films are casted following the solvent evaporation protocols developed for amphiphilic copolymers[33,34]. Fig. 10a displays a picture of a macroscopic film obtained from PANa(5K)-*b*-PAM(60K)/YHA. With a thickness evaluated around 100 μm, the film is transparent and brittle. It shows the same mechanical properties than a poly(acrylamide) film made in the same conditions. We have checked that without addition of polymers, the YHA nanoparticles alone do not form a film, but instead a white powder (Fig. 10b). In order to check the structure of the composite films, the PANa-*b*-PAM/YHA thick films were investigated by SAXS. As illustrated in Fig. 11, the data of a solid state sample exhibits a correlation peak around 0.02 Å$^{-1}$ associated to a *d*-spacing of ~ 320 Å. This peak results from the correlation between the YHA clusters which have sustained the solvent evaporation process. The YHA cores can now be considered as dispersed inclusions in a solid PAM matrix. Also displayed in the figure is the result of an experiment with a PAM film without added particle. In this case, the scattering is flat and no structure peak is visible. In Fig. 11, the scattering cross-section obtained for the thick film compares very well with the form factor of the nanoparticles, especially at large wave-vector. The superimposition observed above 0.2 Å$^{-1}$, *i.e.* on a local scale confirms the presence of the nanoparticles in the film. Finally, we have included in Fig. 6 the $q_{Max}$-values observed on several films, that is for c → 1. We have found systematically $q_{Max}$(c→1) between 0.015 and 0.02 Å$^{-1}$, instead of the 0.025 Å$^{-1}$ extrapolated from the dilute solution scaling. This deviation could be explained by a partial reorganization of the cores during the late stages of the solvent evaporation. In conclusion, we confirm that the nanocomposite thick films achieved by this technique contain 20 % by weight of nanoparticles and ~ 70 - 80 % of polymers (depending on the amount of unbound water molecules).

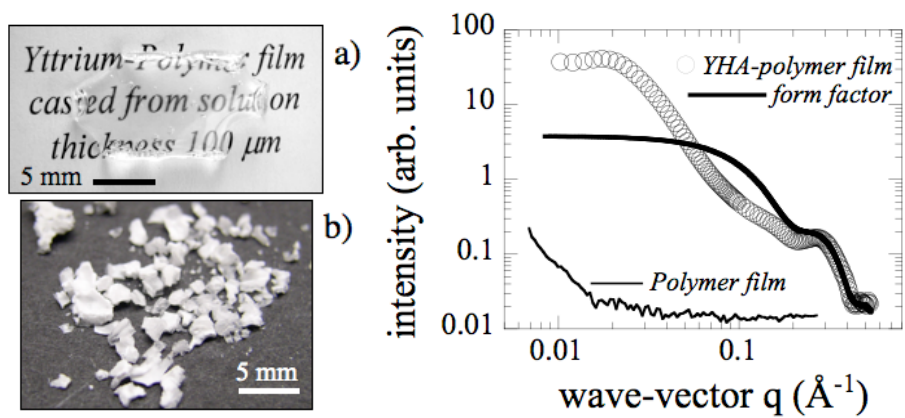

**Figure 10 :** a) Image of a PANa(5K)-*b*-PAM(60K)/YHA thick film obtained by casting. The film is 100 µm thick, transparent and brittle. b) Powder obtained in the same casting conditions as in a) and starting from a YHA-solution with no polymer added.

**Figure 11 :** Comparison between SAXS intensities obtained from a composite film (open symbols), from a YHA-nanoparticle solution (form factor, thick line) and from a poly(acrylamide) film (thin line). The structure peak observed for the composite film at q ~ 0.02 Å$^{-1}$ results from the correlation between the YHA clusters. Note the superimposition observed above 0.2 Å$^{-1}$ between the film intensity and the particle form factor.

# IV – Conclusion

In the present paper, we have shown that the inherent instability of inorganic nanoparticle sols in general may be resolved by complexation with oppositely charged ion containing polymers. In addition to our previous contributions on nanoparticle-polymer self-assemblies[18,21], here we investigate a new property for these aggregates, namely their stability as function of the concentration and their propensity to build nanocomposite films. We have found that the mixed aggregates can sustain the solvent removal up to the solid state. Solid films with thicknesses up to several hundreds of microns were cast from solutions and the structure of the complexes remains unchanged during film casting.

In the dilute regime, x-ray and neutron scattering have revealed the form factor of the supermicellar aggregates. The core-shell microstructure of the supermicellar aggregates was derived from a combination of the scattering (light, x-ray, neutron) and electron microscopy experiments, as well as from a quantitative determination of the different scattering length densities. In order to fit the SANS data in the dilute regime, we used well-established phenomenological equations derived for polymer stars[40,44,45]. In the intermediate regime of concentration, we have observed around c = 10 wt. % viscous liquids, and at c = 40 wt. % viscoelastic gels. The gel-like properties of the latter solutions result from the high packing fraction of the supermicellar aggregates and from the deformation of the neutral brush (increasing then the contribution to the elastic modulus). On the whole concentration range, that is between 1 to ~ 90 wt. %, the position of the primary structure peak revealed in the scattering intensities scales with the total concentration according to $c^{1/3}$, which is characteristic for aggregates with fixed aggregation number and local spherical symmetry. From their core-shell microstructure as well as from their concentration behaviors, the YHA-copolymer mixed micelles bear strong similarities with the amphiphilic block copolymers in aqueous solutions[30].

**Acknowledgements** : We thank Yoann Lalatonne, Julian Oberdisse, Amit Sehgal for many useful discussions. Mathias Destarac, Annie Vacher and Marc Airiau from the Centre de Recherches d'Aubervilliers (Rhodia, France) are acknowledged for providing us with polymers and for the Cryo-TEM images of the mixed aggregates. This research is supported by Rhodia and by the Centre de la Recherche Scientifique in France.

# Tables

**Table I : Chemical formula, molecular weight ($M_W$), molar volume ($V_{mol}$), coherent neutron and x-ray scattering length densities ($\rho_{N,X}$) for the species studied in this work.**

| species | chemical formula | $M_W$ g·mol$^{-1}$ | $V_{mol}$ cm$^3$·mol$^{-1}$ | $\rho_N$ 10$^{10}$ cm$^{-2}$ | $\rho_X$ 10$^{10}$ cm$^{-2}$ |
|---|---|---|---|---|---|
| acrylic acid | CH$_2$CH-COOH | 72.06 | 47.8 | + 2.09 | 13.4 |
| sodium acrylate | CH$_2$CH-COO$^-$,Na$^+$ | 94.04 | 33 | + 4.37 | 24.5 |
| acrylamide in H$_2$O | CH$_2$CH-CONH$_2$ | 71.08 | 53.3 | + 1.86 | 12.0 |
| acrylamide in D$_2$O | CH$_2$CH-COND$_2$ | 73.08 | 53.3 | + 4.19 | 12.0 |
| yttrium hydroxyacetate | Y(OH)$_{1.7}$(CH$_3$COO)$_{1.3}$ | 194.58 | 50.0 | + 3.50 | 31.9 |
| hydrogenated water | H$_2$O | 18.02 | 18.02 | − 0.56 | 9.4 |
| deuterated water | D$_2$O | 20.02 | 18.02 | + 6.40 | 9.4 |

**Table II : Characteristic sizes for poly(acrylic acid)-*b*-poly(acrylamide) block copolymers, yttrium hydroxyacetate nanoparticles and mixed polymer-nanoparticle aggregates.**

| colloidal species | $R_H$ (nm) | $R_G$ (nm) |
|---|---|---|
| PANa(5K)-*b*-PAM(30K) | 5.5 (*l*) | n.d. |
| PANa(5K)-*b*-PAM(60K) | 7.9 (*l*) | |
| YHA-nanoparticle | ~ 2 (*l*) | 1.82 (*n*) <br> 1.65 (*x*) |
| mixed aggregate in : Fig. 4 | 34 (*l*) | 17.2 (*l*); 11.0 (*x*) |
| Fig. 8 | 40 (*l*) | 22.7 (*n*) |
| Fig. 9 | 36 (*l*) | 18.2 (*n*) |